\documentclass[aps,preprintnumbers,nofootinbib,superscriptaddress,twocolumn]{revtex4}

\usepackage{amsmath}
\usepackage{amsthm}
\usepackage{amssymb}
\usepackage{amstext}
\usepackage{graphicx}
\usepackage{url}
\usepackage{multirow}

\theoremstyle{plain}

\theoremstyle{definition}

\newcommand*{\comment}[1]{}

\begin{document}
\title{Might quantum-induced deviations from the Einstein equations
detectably affect gravitational wave propagation?}
\author{Adrian \surname{Kent}}
\email[]{a.p.a.kent@damtp.cam.ac.uk}
\affiliation{Centre for Quantum Information and Foundations, DAMTP, Centre for Mathematical Sciences, University of Cambridge, Wilberforce Road, Cambridge, CB3 0WA, U.K.}
\affiliation{Perimeter Institute for Theoretical Physics, 31 Caroline Street North, Waterloo, ON N2L 2Y5, Canada.}
\date{October 2012 (revised March 2013)}

\begin{abstract}

A quantum measurement-like event can produce any of a number of 
macroscopically distinct results, with corresponding macroscopically
distinct gravitational fields, from the same initial
state.   Hence the probabilistically evolving large-scale structure of
space-time is not precisely or even always approximately
described by the deterministic Einstein 
equations.  

Since the standard treatment of gravitational wave propagation
assumes the validity of the Einstein equations, it is questionable
whether we should expect all its predictions to be empirically
verified.  `In particular, one might expect the stochasticity
of amplified quantum indeterminacy to cause 
coherent gravitational wave signals to
decay faster than standard predictions suggest. 
This need not imply that the radiated energy flux from
gravitational wave sources differs from standard theoretical
predictions.   
An underappreciated bonus of gravitational wave
astronomy is that either detecting or failing to 
detect predicted gravitational wave signals would 
constrain the form of the semi-classical theory of gravity that
we presently lack.  

\end{abstract}


\maketitle

\section{Introduction} 

One difficulty theoretical physicists currently face is  
that, as the subject has grown larger and speculative attempts to 
address fundamental problems have multiplied, our collective 
knowledge has become increasingly fragmented.  Questions which
are at the forefront of the attention of one group of people can be
pretty much neglected, or not even recognised, by others.   
Indeed, even individuals may display a version of this: 
because our attention is selective and trained, we can end 
up functioning according to a sort of doublethink, in which we
note a problem in one context and neglect it in others.  
I suspect this is actually much more common than we generally
realise. 

Here, following a broader
programme (\cite{akcausalqt,aklocalcausality,salart,bgqt,satellite}; see also
\cite{penrose,diosi,pearlesquires} for earlier distinct but related 
ideas in this area) of trying to
test underexplored foundational
questions relating quantum theory and gravity, I suggest one area --- gravitational wave physics ---
where this phenomenon may be at work. 
The problem is this.  On the one
hand, the standard theory of propagating gravitational waves
treats them as perturbations of the Einstein gravitational
field equations.  On the other hand, there is no way of 
incorporating unpredictable quantum events into a classical
theory of gravity described by the Einstein equations.     

This is not only because of small corrections arising from 
an as yet unknown quantum theory of gravity.  If this were the
only issue, it might be easier to defend the case that the 
standard perturbative treatment of gravitational waves is 
likely to be essentially
unaltered by quantum corrections, at least in regions where
the gravitational field is not too strong.  
The more immediate problem is that, whenever
a quantum measurement type interaction takes place --- whether a deliberate
measurement by a human observer or a naturally occurring event --- 
it can produce any of a number of 
macroscopically distinct results from the same initial
state.  According to the standard understanding of quantum
theory, these measurement outcomes are intrinsically probabilistic.
Not only can they not be predicted in advance by quantum theory, but 
there are very strong reasons \cite{bellbook,chsh,cr,pbr} to believe that no underlying 
deterministic theory allows us to predict them.  
These macroscopically distinct outcomes lead to 
distinct space-times and matter distributions.  Each of several possible
space-times and matter distributions could thus emerge  from the same
initial state.   If the measurement event is suitably amplified,
their differences can be arbitrarily large.   
Since general relativity is deterministic, it 
follows that the Einstein equations cannot even approximately
describe the large-scale structure of space-time around measurement
events.  

If we try to describe the space-time physics in the vicinity of the
measurement event by some classical stress-energy tensor and metric, it seems
that we need to introduce some stochastic source which creates
an unpredictable macroscopic perturbation of the metric and tensor, at
some point -- or, probably more accurately, in some region -- in the 
vicinity of the event.
However, not only are the Einstein equations inconsistent with these
perturbations, but moreover we know of no generally valid way of
incorporating them into some semi-classical theory coupling the metric
to a classical quantity derived from the quantum stress-energy tensor,
or by some stochastic modification of the Einstein equations.  

Looking at current knowledge without a predefined theoretical
agenda, then, all we can really say for sure is that general relativity and quantum theory
both work well in their respective domains.   However, even characterizing precisely those 
domains of confirmed validity is subtler and harder than it first
seems, as the comments above illustrate.
It is even harder to justify confidence that we now understand all the
principles underlying a unified theory.   It's not evident that
mainstream ideas work, and while it's certainly not evident that
comparatively undeveloped alternatives will work either, they do exist.  

For example, it is certainly possible to imagine unifications
in which both general relativity and quantum theory work as good
approximations in their respective domains, and in which gravity
is quantized, but nonetheless the structure of space-time is also 
constrained by additional laws that modify the probability of 
each space-time and are defined by intrinsically geometric rules
that do not follow from any quantum theory \cite{bgqt}.   
It's also possible to imagine theories in which gravity is 
not quantized at all, and the laws of nature define some
probability distribution on classical space-times with 
quantum matter distributions \cite{bgqt}.   

The general hypothesis that gravity and quantum state
collapse could be linked \cite{penrose,diosi,pearlesquires}, via
fundamentally non-unitary dynamical laws is also
intriguing, even though it too is hard to make into a precise theory.   

Also, even if one of the more currently popular approaches to quantum gravity {\it is} correct,
it is very unclear how to derive from it a phenomenological higher-level theory 
that couples microscopic quantum matter with macroscopic events in
space-time, or precisely what features such a theory would have. 

To reiterate, the point here is not to advocate these comparatively 
undeveloped alternative ideas, but simply to underline that unifying
gravity and quantum theory is an open subject and there are 
many un(der)explored possibilities.   
Even in macroscopic regimes with weak gravitational fields,  
we simply do not have a theory of matter
and space-time good enough to fit all observable data.   
This problem occupies a peculiar status in modern physics: it cannot
be denied that the problem is there, but yet most discussions of
quantum theory and gravity ignore it.   The gap in the literature
is so glaring that one almost gets the impression
that it is somehow seen as scientifically unsophisticated to look for a theory -- even a provisional
phenomenological theory -- that actually fits the available empirical data.   

To sum up: {\it the Einstein equations do not actually correctly describe the
large-scale structure of space-time and finding equations that do 
is an open question}. It thus doesn't seem so 
obvious that the long-range propagation of gravitational waves is
necessarily correctly described by considering them as perturbations of the
Einstein equations.

The rest of this paper attempts to flesh out
this point mostly by conceptual, rather than mathematical, 
argument.  In mitigation, I would stress again that we know
nothing for certain about unifying quantum 
theory and gravity, and the world possibly already has more than enough 
mathematically rigorous, but conceptually problematic, and quite likely
ultimately irrelevant, calculations based on 
speculative mathematical ideas about how to solve the problem.  

Our ultimate aim, of course, is to 
test precise mathematical theories against quantitative experimental
data, but just at the moment we need new ideas about where
to look.  It seems to me there are strong reasons to try a different style
of scientific reasoning: namely, to look at interesting experiments and
observations where we don't yet know {\it for sure} what
theory predicts but can -- now, or soon -- get an empirical answer,
and to ask whether there is any semi-plausible phenomenological model
or intuition that might contradict the standard expectation (if there is
one).   Either we verify with certainty interesting features of 
gravitational physics that were previously either ignored completely or
assumed without compelling evidence, or (even better) we learn something new
and surprising.    

\section{Prior tests of probabilistic semi-classical gravity} 

\subsection{The Page-Geilker Experiment} 

The fundamental problem in constructing semi-classical gravity
theories was illustrated by a very simple
experiment carried out some time ago by 
Page and Geilker \cite{pagegeilker}.
Page and Geilker's aim was to refute conclusively the 
hypothesis that a classical gravitational field couples to
quantum matter via the semiclassical Einstein equations
\begin{equation}\label{semiclassical}
G_{\mu \nu} = 8 \pi \langle \psi \, | \, T_{\mu \nu} \, | \, \psi \rangle \, ,
\end{equation} 
where $G_{\mu \nu}$ is the Einstein tensor and $T_{\mu \nu}$ the 
quantum stress-energy operator.  Their  hypothesis presupposes an
Everettian interpretation of quantum theory, so that the matter field
quantum state, $ | \, \psi \rangle$, is supposed to evolve
unitarily without collapse.

As Page and Geilker noted, this hypothesis already seemed intrinsically
unlikely (perhaps even incredible) before they
carried out their experiment, since it is hard to imagine 
how it could lead to a cosmological theory which accounts for
our observation of gravitational fields generally consistent with 
those predicted by the Einstein equations from
the observed positions of astronomical bodies. 
Page and Geilker's motivation for their experiment is thus seriously 
questionable.  Perhaps, though, one {\it could} imagine some form of theory in 
which a classical gravitational field couples to the expectation
value of quantum matter, while the quantum matter state over time collapses towards 
definite values for $T_{\mu \nu}$.  
In any case, it still seems good to have conclusive experimental confirmation
even of very solidly based theoretical expectations when, as here,
we have no complete theory.  

The experiment proceeded by counting the number of decays detected 
from a radioactive source in a given time interval, and then 
manually placing large ($\approx 1.5$ kg) lead balls in one
of two configurations, with the choice of configuration
depending on the decay count.   A Cavendish torsion balance,
sensitive enough to distinguish between the two configurations,
was used to measure the local gravitational field.   
To good approximation, Eq. (\ref{semiclassical}) predicts that 
in each run the experimenter should (whichever of the two
configurations they place the masses in) observe a gravitational
field defined by the weighted average of the fields corresponding to
the two possible configurations.  
As most expected, the results were consistent with the hypothesis
that the gravitational field is determined by the 
configuration of the masses chosen in any given run of
the experiment, and inconsistent with Eq. (\ref{semiclassical}).

We can flesh out the implications of the experiment --- or, to capture
the historical flow of ideas better, 
perhaps one should say
the implications of the generally held prior intuition that its results
would be those which were in fact observed --- by looking at three possible
solutions to the Einstein equations.   

First, consider
the classical metric and matter fields, which we denote
by $( g^0_{\mu \nu} , \phi^0 )$, in the neighbourhood of
some spacelike hypersurface $S$ before the point at which a Page-Geilker
experiment is carried out.   By ``the classical metric and matter
fields'', we mean here the fields that would ordinarily be defined by someone 
trying to model the local space-time neighbourhood using general
relativity --- for instance, an engineer, trying to predict how small
lumps of matter will evolve, and doing so as precisely as is possible 
without taking quantum theory into account.  
Let us suppose there is some way
of fitting these data to a solution of the Einstein equations,
using some well-defined and natural recipe (not necessarily 
Eq. (\ref{semiclassical})) to obtain a classical stress-energy
tensor from the quantum matter field, and denote the corresponding
spacetime by $\Sigma_0$.  

Now consider the classical 
metric and matter fields, $( g^1_{\mu \nu } , \phi^1 )$
and $( g^2_{\mu \nu } , \phi^2 )$, in the neighbourhood of a spacelike 
hypersurface $S'$ after the point at which a Page-Geilker
experiment is carried out.
Let us suppose these data can also be fitted to solutions of 
the Einstein equations,
using the same recipe for a classical stress-energy
tensor as before, and denote the corresponding
spacetimes by $\Sigma_1$ and $\Sigma_2$.  
 
Since $\Sigma_1$ and $\Sigma_2$ describe macroscopically distinct
matter configurations on $S'$, they are not identical, and so
cannot both be identical to $\Sigma_0$.  In fact, since neither of them is
preferred in any way, one would not generally expect {\it either} of them
to be identical to $\Sigma_0$, assuming that the recipe used to 
obtain the stress-energy tensors depends in any natural way on
the relevant fields.  (Obviously, if completely arbitrary recipes
are allowed, one could contrive things so that
one of them equals $\Sigma_0$.  For instance, one could 
define the recipe for obtaining
$\Sigma_0$ to involve first studying the possible experimental outcomes, 
then constructing $\Sigma_1$ and $\Sigma_2$, and then simply setting $\Sigma_0$
to be equal to one of them.)     
  
To sum up, then, if we can find a way of describing 
the data before and after the experiment by piecewise 
continuous solutions of the classical Einstein equations, 
they will generally be solutions that form part of 
different spacetimes.  In this description (if there is
indeed such a description), it is as though some sort of stochastic
jump takes place, starting from one solution, and arriving
at one of two alternative solutions, both distinct
from the original.  

Is there such a description?  Can the spacetime
we actually observe be described by piecewise continuous
solutions of the Einstein equations?   I don't think
we know for sure: some sort of smoothing could take 
place in the vicinity of Page-Geilker experiments, for
instance.  But it seems to be generally tacitly assumed --- and 
it seems to be necessary to assume, in order to explain
experimental data --- that this is 
at least approximately the case.  For, on the one hand,
if it were not the case that large regions of spacetime 
are well described by
a solution to the Einstein equations, we would not be
able to account theoretically for any of the confirmed predictions of
general relativity.   On the other hand, as we have just argued,
the sort of macroscopic indeterminacy exhibited by the Page-Geilker
experiment implies that spacetime cannot be described globally
by a single solution of the Einstein equations.  

\subsection{Probing gravitational non-locality and the Salart et al. experiment}

A more recent proposal \cite{aklocalcausality} with some related motivations was to look for direct
evidence of violations of Bell's local causality in the gravitational
field.   Recall that Bell experiments (modulo loopholes) show that any hidden variable theory underlying
quantum theory must violate Bell's condition of local causality.
Since non-local hidden variable theories are theoretically
uncompelling and difficult to reconcile with relativity, this  
gives much stronger evidence that the outcomes of quantum experiments
are indeed inherently unpredictable.    
The evidence that the classical gravitational field evolves
probabilistically, though, is less direct. 
While the Page-Geilker experiment {\it appears} to confirm that the
gravitational field evolves probabilistically, one could still imagine an
underlying deterministic semi-classical gravity theory, sensitive to microscopic variables,
that predicts each observed evolution, in the same way that 
de Broglie-Bohm theory and other deterministic hidden variable theories 
reproduce the predictions of quantum theory.  
We would like to be able to argue directly that any  
deterministic phenomenological theory of gravity {\it must} be
non-local.   Since a non-local gravity theory would be very hard to reconcile with either special or general
relativity, this would be a compelling reason for abandoning all hope for
deterministic semi-classical theories.  

Another strong motivation for verifying this point directly
is that gravitational collapse models highlight another possible loophole
in the interpretation of all Bell experiments to date.
This ``collapse locality loophole'' \cite{akcausalqt} arises because collapse models
suggest that a definite measurement outcome occurs only after 
macroscopic amplification to a particular scale (which depends
on the parameters of the collapse model).  If this is correct,
to exclude locally causal explanations we need Bell experiments
that ensure that collapses, and thus definite measurement outcomes,
take place in spacelike separated regions in the two wings.  
{\it No Bell experiment to date ensures this for the full range of collapse model
parameters consistent with known experiment} \cite{akcausalqt}.

This motivation is further reinforced by the observation \cite{akcausalqt} that there {\it are} 
ways in which a consistent theory combining quantum theory and gravity could
conceivably produce the outcomes observed in all Bell experiments to
date and nonetheless allow only locally causal evolutions of the
metric.  Of course, models incorporating these ideas have highly
non-standard properties, and may be theoretically problematic as 
well as ad hoc.   Still, as with the Page-Geilker experiment, clear experimental data
would be much preferable to strong theoretical intuitions and arguments.   

A beautiful experiment investigating this possibility was
carried out by Salart et al. \cite{salart}, showing that the local causality
loophole can indeed be closed at least for gravitational collapse models
whose collapse times and scales agree with theoretically 
motivated estimates proposed by Penrose \cite{penrose} and Diosi \cite{diosi}.   
Further more conclusive experiments have been proposed \cite{satellite}, with the
ultimate aim of directly measuring non-locally correlated
gravitational fields in such a way that these measurements are
themselves completed in space-like separated regions.  

\section{Natural measurement-like events}

Of course, quantum measurement-like events with macroscopically
distinct consequences take place without any artificial help.    
On Earth, fissioned particles and cosmic rays leave tracks
in mica; frogs can respond to the stimulus
of a single photon; a single gamma ray or charged
particle can trigger a cancer; the bouncing of photons
and cosmic rays off dust particles must from time to 
time determine the formation or otherwise of a 
particular macroscopic clumping.    Each of these
outcomes is effectively a quantum measurement of 
the position of a particle whose wave function was
previously delocalised.   
On the cosmological scale, quantum fluctuations are
believed to have seeded the instabilities that led
to galaxy formation.  
 
It would be very interesting indeed to try to 
quantify the degree to which quantum noise, 
bubbling up from the microscopic realm, affects
the predictability of the macroscopic world in
general, and in particular to characterise the degree and type of the 
resulting deviations from Einstein's equations.   
This project is beyond our present scope, though.  

For the purposes of the present discussion, we need only take the point
that these deviations occur naturally, and presumably
have been doing so since very early cosmological times.  
This is why the outcome of the Page-Geilker experiment
was generally (perhaps even universally) anticipated.
In other words, while the Page-Geilker experiment is a good illustration of 
the point that our observed space-time deviated from 
Einstein's equations, we do not actually need to appeal to
it to make that point.  

\section{Quantum gravity: resolution or distraction?}

According to one school of thought, at this point in the
discussion one should throw up one's hands, regret the
fact that we don't yet have a quantum theory of gravity, 
and accept that we can't productively advance the 
discussion further without one.  

It seems to me
far from evident that we should heed this
counsel of despair.  I can see
two reasons for optimism.

First, we might not actually need a 
quantum theory of gravity at all.
Second, even if we do, it ought to imply some effective
phenomenological theory of classical gravity which
incorporates stochastic fluctuations into general
relativity.  
In the first case, we might hope for
some {\it fundamental} theory which incorporates
stochastic fluctuations into general relativity;
alternatively, we might hope for a theory based on
different principles, which again should imply
an effective phenomenological theory of classical gravity
of the type just mentioned. 

In all these cases, it is reasonable to try to explore how
a classical gravity theory with stochastic fluctuations 
might be probed experimentally. 
But we don't have such a theory.  Perhaps the best hope,
then, is that experiment might guide us to the right theory, if we 
can at least identify what to look for experimentally.  

The following two sub-sections flesh out these arguments in more detail. 

\subsection{Do we need quantum gravity to explain the Page-Geilker experiment?}

Embarrassingly, our best theory of gravity, general relativity,
has no way of consistently incorporating the results of 
macroscopically amplified quantum measurement events, 
whether they occur naturally or are created artificially
as in a Page-Geilker experiment.  
The current conventional wisdom suggests that 
this embarrassment stems from our failure to 
devise a consistent quantum theory of gravity.   

There is --- the standard argument runs --- no
fully consistent way to couple a classical gravitational field 
to quantum matter fields: the gravitational field also 
needs to be quantised.  
We would expect --- the argument proceeds to suggest --- 
that in a full quantum theory of gravity, the gravitational field  
would evolve so as to be (to very good approximation) correlated 
with the matter fields in any given branch of the 
universal wave function.  In particular, we would expect
a full quantum theory of gravity to predict the observed
outcome of the Page-Geilker experiment: this is why Page
and Geilker provocatively titled their paper ``Indirect Evidence for
Quantum Gravity''.  More generally, we would expect a 
full quantum theory of gravity 
to predict that the gravitational field should be correlated
with the positions of astronomical bodies in the way 
we observe.   
    
Of course, this argument begs several key questions. 
Aside from the problems of principle in unifying quantum
theory and gravity, discussed above, there is the  
problem of making scientific sense of purely unitary quantum theory.
We do not have, despite nearly fifty years' of effort,
any clearly consistent and logically compelling account 
of how Everett's original intuition might be fleshed out
into a clearly and carefully justified interpretation of a unitarily evolving 
universal wave function.  (State-of-the-art reports and assessments of recent
attempts can be found in Ref. \cite{mwbook}.) 

\subsection{Probing an effective theory derivable from quantum gravity}

What if some version of quantum gravity is correct, though?   
Suppose, for example, we find some rigorously defined way 
of carrying out path integrals over gravitational and matter field
configurations, and find some evidence that it gives correct
answers.  In order to understand large-scale gravitational physics, 
we would {\it still} need some (presumably) phenomenological effective
theory, derived from our fundamental quantum gravity theory, which
characterises the quasiclassical behaviour of matter and gravity
that we actually observe.  
(In Gell-Mann and Hartle's terminology \cite{gmh}, we would need some way
of characterising our own quasiclassical domain within this
hypothetical quantum gravitational or quantum cosmological
theory.)  

In particular, this higher-order theory would need
to be consistent with the Page-Geilker experiment
and with the observed correlations of gravitational fields 
and astronomical bodies.   It thus seems a
reasonable conjecture --- suggested by observational evidence, 
and contradicted by nothing we 
know about quantum gravity --- that we would end up with some sort
of stochastically modified version of general relativity, albeit 
in this case as a derived effective theory rather than a fundamental
theory.  If so, one might make the further reasonable-looking guess that the propagation
of gravitational waves is approximately described by considering them
as perturbations of the gravitational field {\it 
within this higher-order
quasiclassical theory}  
 
\section{What happens to gravitational waves in a stochastic
  modification of general relativity?} 
  
Without knowing the details, one can only guess. 
So, without further ado, I shall.  A plausible guess, it
seems to me, is that stochastic fluctuations break
up the coherence of propagating waves.  It is 
difficult to hear someone shouting in a high wind, not
only because the noise of the wind drowns out the 
propagating sound wave, but also because the turbulence
causes its amplitude to decay faster than in still air.  

If the level of stochastic fluctuations is constant throughout 
a region in which a wave propagates, the simplest guess
would be that the wave amplitude decays by a factor exponential in
the region length, in addition to the normal approximately inverse
square law decay.   Without knowing the theory, one can't 
estimate the value of the exponential constant --- but if this
guess is right, and if gravitational wave astronomy turns out
nonetheless to be viable, one might be able to estimate it
from observational data, and thereby get quantitative data
characterising an important feature of the relation between
quantum theory and gravity.   

This raises the possibility that the stochastically induced
decay of gravitational waves could conceivably prevent
gravitational wave astronomy from being viable with
presently envisaged gravitational wave detectors.
If so, of course, gravitational wave astronomy's loss would be 
gravitational theory's gain.   

\section{What about the binary pulsar observations?}

If one suggests
the possibility that the standard account of
gravitational wave physics might not be correct, 
one has to deal with the counter-argument that observations of binary
pulsars \cite{binarypulsar} have already confirmed the standard account to a 
very impressive degree of precision. 
This counter-argument has no force against the speculations
considered here, though.   The suggestion is not that binary
pulsars do not emit gravitational waves, and thereby lose energy,
as standard theory predicts.  The suggestion is, rather, that 
the gravitational waves lose coherence, and thus decay faster
than expected, as they propagate through space, and hence in
particular that gravitational wave signals reaching Earth might
be weaker than anticipated.   Careful observation of a drum
vibrating in the distance would reveal that it is losing
energy by radiating sound waves; nonetheless, the sound of
the drum will not propagate as far in a strong wind.  
There is no evident inconsistency here.   

\section{Comparing Quasiclassical Gravity and 
  Quasiclassical Electrodynamics}

To what extent are the problems we raise about our understanding
of quasiclassical physics specific to gravity?    
In particular, are there any reasons to think that classical gravitational waves
might behave differently from classical electromagnetic waves?  

In considering these questions, it's helpful first to consider 
quasiclassical electrodynamics in Minkowski space.
Clearly, some of the points made above apply.
In particular, we can carry out Bell experiments and ensure
that, on each wing, a source of electromagnetic waves behaves differently 
depending on the measurement setting and outcome on that wing, and
that
the measurement settings themselves are locally determined by random
quantum events.   For
example, a charged sphere could be move in any of four different ways,
depending on the two measurement choices and two possible outcomes,
and the measurement choices could be determined, just before the
measurements are made, using bits produced by quantum random number generators.   
Since the outcomes of Bell experiments are nonlocally correlated,
we expect this to produce nonlocal correlations in the electromagnetic
fields propagating from the regions of the two measurements. 

Now, this probably has not been directly tested in experiments to
date, and I am not sure we can in principle rigorously exclude models
(with very counterintuitive features) that agree with experiments to
date but predict that nonlocal correlations of classical
electromagnetic fields cannot in fact be observed.  
Of course, this would be a very surprising outcome indeed.
We ignore the possibility here, since our aim is to understand
whether one might have possible reasons to look for unexpected
behaviour in quasiclassical gravitational physics even if 
there are no analogous surprises in quasiclassical electrodynamics. 

Assuming, then, that nonlocal correlations can be created in 
macroscopic electromagnetic fields, it follows that 
quasiclassical electrodynamics in the real world cannot be described by an underlying
local deterministic model.  Note, though, that the nonlocalities we introduced 
arise entirely from nonlocal correlations in the motion of sources.   Given a
description of the
motion of each source, we can calculate the subsequent behaviour of
the electromagnetic fields it generates.  Since electrodynamics is a linear theory,
we can obtain a complete solution by superposing the contributions from the 
various sources.   This gives a strategy for building a
phenomenological model of quasiclassical electrodynamics in the
presence of quantum unpredictability and nonlocality: first apply the
predictions of quantum theory to give a model of additional
stochastic (and nonlocally correlated) forces acting on the sources,
and then solve to obtain the fields.   
Adding forces that alter the motion of the sources does not affect
charge conservation, so in such a model we still have $\partial_{\mu}
J_{\mu} = 0$.   

It would be wrong to suggest this gives a rigorous understanding
of the relationship between quantum and quasiclassical electrodynamics
in Minkowski space.  We do not even have a completely rigorous
definition of quantum electrodynamics as a non-trivial theory.
Nor do we have a precise general prescription for how to obtain
quasiclassical equations of motion from quantum theory, either for
electrodynamics or for any other physically relevant theory. 
However, we {\it do} at least have an ansatz for dealing with the
quasiclassical consequences of quantum experiments with unpredictable and
nonlocally correlated outcomes, and this ansatz does not violate the
conservation laws necessary for a consistent solution of the
electrodynamic equations.   

Now compare the situation when we try to model an analogous experiment  
in which the measurement choices and outcomes of Bell experiments 
determine the motion of massive objects on each wing, with the
measurement choices again locally determined by quantum random
number generators.   As noted earlier, we cannot model the 
quasiclassical physics by extrapolating the predictions of general
relativity from data on a spacelike hypersurface before the Bell experiment, since general
relativity is deterministic and the Bell data are not.  
So far the analogy with electrodynamics holds, since electrodynamics is also
deterministic.  
However, we run into further problems in this case. 
To define any consistent solution of the Einstein equations,
we need the local conservation of stress-energy, $D_{\nu} T^{\mu \nu}
= 0$.   We know of no generally covariant quasiclassical model of 
the possible outcomes of quantum measurement-like processes that
preserves stress-energy and is consistent with general relativity
where quantum effects are negligible.  
(Indeed, even non-relativistic dynamical collapse models
\cite{ghirardi1986unified,GPR}, which might be the best 
guesses at phenomenological descriptions of the quasiclassical
physics emerging from measurement interactions, violate conservation
of energy.)  Without such a model, it seems our best 
description of quasiclassical gravitational physics would be by
models which generally obey 
the Einstein equations but have singularities or discontinuities
at or in the vicinity of quantum measurement events. 
And if that {\it were} the best possible description, the  
standard classical derivation of gravitational wave propagation would 
break down in these regions.    

To be sure, there are further uncertainties here.   If these 
discontinuities are physically real, should we  
expect them to affect the propagation of electromagnetic radiation 
in the same way as they affect the propagation of gravitational
waves?   If so, of course, any effect is likelier to be 
evident in standard (electromagnetic wave observation) astronomy
than in gravitational wave astronomy, and the absence of any
observed effect to date is a strong constraint.  
On the other hand, we have a quantum theory of electromagnetism
and no quantum theory of gravity.   And, if there {\it is} a 
quantum theory of gravity from which quasiclassical solutions
obeying Einstein's equations with discontinuities emerges, we 
have no clear reason to think that coherent beams of gravitons and photons 
should scatter similarly from the discontinuities -- indeed one might
guess that gravitons are more directly affected than photons by a
discontinuity in the classical field generated by gravitons.  

Some may nonetheless hold the intuition that we should expect exactly the
same effects in quasiclassical gravity and quasiclassical
electrodynamics.   The points made here do not refute this
possibility, but they do give significant reasons to query it.   

\section{Summary}

In this paper, we raised a possibility that does not seem
to have been considered: that stochastic corrections to the
Einstein equations dissipate gravitational waves.  Such stochastic
corrections could either arise directly from a fundamental theory or as
a phenomenological effect resulting from quantum gravity (or some
other presently unknown type of theory).  Either way, our guess
at their effect on gravitational wave propagation is not provable
given the present state of theoretical understanding.   But is it obviously wrong, or totally implausible?  
If, as we suggest, not, it seems a possibility to be kept in mind
if and when gravitational wave astronomy produces 
data, null or otherwise.  
We hope too that the questions raised here may encourage more
attention to be focussed on 
the problem of finding realistic quasiclassical descriptions
of gravitational physics in the presence of quantum measurements,
through Bell experiments and otherwise.   

\section{Acknowledgements}
This work was partially supported by a Leverhulme Research Fellowship, a grant
from the John Templeton Foundation, and by Perimeter Institute for Theoretical
Physics. Research at Perimeter Institute is supported by the Government of Canada through Industry Canada and
by the Province of Ontario through the Ministry of Research and Innovation.

\end{document}